\documentclass[aps,prb,showpacs,reprint]{revtex4-1}

\usepackage{graphicx}
\usepackage{amsmath}
\usepackage{bbm}

\begin{document}
\title
{Quantum size effect on the paramagnetic critical field in Pb nanofilms}
\author{P. W\'ojcik}
\email[Electronic address: ]{pawelwojcik@fis.agh.edu.pl}
\affiliation{AGH University of Science and Technology, Faculty of
Physics and Applied Computer Science, al. Mickiewicza 30,
Krak\'ow, Poland}
\author{M. Zegrodnik}
\affiliation{AGH University of Science and Technology, Academic Centre for Materials and Nanotechnology,
al. A. Mickiewicza 30, Krak\'ow, Poland}

\begin{abstract}
The quantum size effect on the in-plane paramagnetic critical field in Pb nanofilms is 
investigated with the use of the spin-generalized Bogolubov-de Gennes equations. 
It is shown that the critical field oscillates as a function of the nanofilm thickness with the period of $\sim2$~ML
(even-odd oscillations) modulated by the beating effect. 
This phenomena is studied in terms of the quantization of the electron energy spectra caused by the confinement 
of the electron motion in the direction perpendicular to the sample. The calculated values of critical fields 
for different nanofilm thicknesses are analyzed in the context of Clogston-Chandrasekhar limit. The influence 
of the thermal effect on the magnetic field induced superconductor to normal metal transition is also discussed. 
Furthermore, the thickness-dependence of the electron-phonon coupling and its influence on 
the value of the critical magnetic field are studied.
\end{abstract}


\maketitle
\section{Introduction}
The interplay between superconductivity and quantum confinement 
has attracted growing interest due to the unique phenomena which appear if the 
electron motion is limited to the size smaller than the coherence 
length.\cite{Zgirski2005,Tian2005,Jankovic2006,Altomare2006,Bezryadin2000,Savolainen2004,Shanenko2006}
The huge progress in nanotechnology which has been made in the last decade allows to 
prepare uniform ultrathin films~\cite{Pfenningstorf2002,Zhang2010,Uchihashi2011,Qin2009,Zhang2010,Uchihashi2011} 
with new superconducting properties which were predicted many years ago.\cite{Blatt1963} 
The oscillations of the superconducting energy gap as a function of the 
nanofilm thickness were predicted by Blatt and Thompson\cite{Blatt1963} in 1963.
The authors of Ref.~\onlinecite{Blatt1963} showed that the quantum-well states which are created 
due to the confinement of the electron motion in the direction perpendicular to the film,
greatly modulate the electron density of states near the Fermi level and thus significantly 
affect the superconducting energy gap of the nanofilm. 
As a result, one would expect a significant increase of the 
energy gap  each time the bottom of a subband passes through the Fermi sphere. 
Nevertheless, either the experimental studies carried out in these years did not exhibit such 
an effect or the observed oscillations differed quantitatively from the theoretical predictions.\cite{Orr1984}
This inconsistency was attributed to the technological  difficulties in the preparation of 
uniform films which were typically polycrystalline and contained a large number of defects. 
Since then, many technological obstacles were overcame what reopened the issue for the quantum 
confinement effect on the superconductivity in the nanoscale regime.

Recently, Guo at al.\cite{Guo2004} have fabricated ultrathin Pb films on Si(111) substrate and 
observed the oscillations of the  critical temperature as a function of the number of atomic 
monolayers. The origin of $T_c$ oscillations associated with the quantum confinement has 
been confirmed independently by the measurements of the quantum-well energies by using photoemission 
spectroscopy. The results presented in Ref.~\onlinecite{Guo2004} have concerned only films with the thickness greater than $20$~ML
because of problems with stability of the even-layered films below $22$~ML.
The study of Pb nanofilms has been extended by Eom et al.\cite{Eom2006} who have reported $T_c$ oscillations 
in the epitaxially grown crystalline Pb films in the thickness range $5-18$~ML.
In Ref.~\onlinecite{Eom2006} the critical temperature has been measured using scanning tunneling micoroscopy
which allows to avoid ambiguities associated with the Au layer which 
has to be applied in the transport measurements reported in Ref.~\onlinecite{Guo2004}. 
It has been found that there is a direct correlation between the oscillations of the 
density of states at the Fermi level and $T_c$ oscillations. 
Also, the measurements of the critical temperature for Pb nanofilms on Si(111) substrate have 
revealed one more interesting phenomena called bilayer or even-odd oscillations. As it has shown, the 
critical temperature for nanofilms with even number of monolayers is greater than the one for nanofilms 
with odd number of monolayers in proper ranges of nanofilm thickness. This dependence reverses 
(odd monolayer films have the critical temperature greater than even monalayer films)
with the periodicity of $7-9$~ML depending on the experiment.\cite{Guo2004,Eom2006}
This effect has been theoretically investigated 
by Shanenko et al. in Ref.~\onlinecite{Shanenko2007} and observed 
in many experiments.\cite{Ozer2006,Ozer2007,Zhang2010}

The new direction of studies devoted to superconductivity in nanoscale 
regime concerns the effect of quantum confinement on the superconductor to normal metal transition
induced by the magnetic field. It has been theoretically shown, that a cascade of jumps 
in the energy gap as a function of the magnetic field occurs in the metallic nanowires.\cite{Shanenko2008} This effect has been 
explained in terms of depairing in the subsequent bands when the magnetic field increases.~\cite{Shanenko2008}
The oscillations of the perpendicular upper critical field in ultrathin lead films have been reported 
by Bao et al. in Ref.~\onlinecite{Bao2005}. This experimental report has shown the even-odd oscillations 
of the upper critical field for Pb nanofilms.
It is also worth mentioning that very recently an interesting effect has been found in amorphous Pb nanofilm.~\cite{Gardner2011} 
Namely, the increase of the critical temperature has been observed after application of a parallel magnetic field to the sample.

According to our knowledge, the theoretical study of  the superconducting-normal metal transition
induced by the in-plane magnetic field and its interplay with  the quantum confinement 
has not been reported until now. In the present paper we investigate the 
superconductor-normal metal transition driven by the in-plane magnetic field for Pb nanofilms. 
We show that the critical field oscillates as a function of the nanofilm thickness 
with the period of $\sim2$~ML (even-odd oscillations) modulated by the beating effect with the 
periodicity of $7$~ML.
The calculated paramagnetic critical field differs from the Clogston-Chandrasekhar paramagnetic bulk 
limit.\cite{Clogston1962,Chandrasekhar1962}
We also discuss the thermal effect and determine formulas for
$H_c(T)$ in the limit $T \rightarrow T_c(0)$.
Finally, we take into account the oscillatory behavior of the electron-phonon coupling occurring in nanofilms
and calculate $H_c$ as a function of the nanofilm thickness for the experimental samples for which a good
agreement of theory and experiment has been found in the study of the critical temperature.~\cite{Chen2013}

The present paper is organized as follows.
In Sec.~\ref{sec:model} we introduce the basic concepts of the calculation
scheme based on the Bogoliubov-de Gennes equations. 
In Sec.~\ref{sec:results} we analyze the results, while the
summary is included in Sec.~\ref{sec:concl}.

\section{Theoretical method}
\label{sec:model}
The superconducting properties of the conventional phonon-mediated pairing system for the high quality
single crystalline structures can be described with the use of the BCS theory, which leads to
the Bogoliubov-de Gennes (BdG) equations in the form
\begin{equation}
\left (
\begin{array}{cccc}
H_e^\uparrow & \Delta & 0 & 0  \\
\Delta ^* & -H^{\downarrow *}_e, & 0 & 0 \\
0 & 0 & H_e,^ \downarrow & \Delta  \\
0 & 0 & \Delta ^* & -H^{\uparrow *}_e
\end{array}
\right ) 
\left (
\begin{array}{c}
\mathcal{U}_i ^\uparrow \\
\mathcal{V}_i ^ \downarrow \\
\mathcal{U}_i ^ \downarrow \\
-\mathcal{V}_i ^ \uparrow
\end{array}
\right )=
E_i
\left (
\begin{array}{c}
\mathcal{U}_i ^ \uparrow \\
\mathcal{V}_i ^ \downarrow \\
\mathcal{U}_i ^ \downarrow \\
-\mathcal{V}_i ^ \uparrow
\end{array}
\right ),
\label{BdG_start}
\end{equation}
where $\mathcal{U}_i ^\sigma$ and $\mathcal{V}_i ^\sigma$ are 
the spin-dependent electron-like and hole-like wave functions ($\sigma=\uparrow, \downarrow$), 
$\Delta$ is the superconducting  energy gap and $E_i$ is the quasi-particle energy. 
In the presence of the in-plane magnetic field $H_{||}$ the Hamiltonian  $H_e ^\sigma$  is given by
\begin{equation}
H_e ^\sigma = \frac{1}{2m} \left ( -i\hbar \nabla -\frac{e}{c} \mathbf{A} \right )^2 + s\mu_B H_{||} - \mu _F,
\end{equation}
where $s=+1$ corresponds to the spin $\sigma=\uparrow$ and $s=-1$ is related to the spin $\sigma=\downarrow$,
$m$ is the free electron mass, $\mu_B$  is the Bohr magneton, $\mu_F$ is the chemical potential and
$\mathbf{A}$ is the vector potential related to the parallel magnetic field. \\	
In the present paper we neglect the orbital effects and consider the superconducting nanofilms in 
the clean paramagnetic limit. This approximation is justified for the 
nanofilms with the thickness less than the magnetic length $a_H=\sqrt{\hbar c / m H_{||}}$. 
The use of the Clogston-Chandrasekhar paramagnetic field for Pb $H^{CC}=13.4$~T gives $a_H=5.6$~nm 
which means that the paramagnetic approximation can be used for Pb nanofilms with thickness 
less than $\sim20$~ML (we assume the lattice constant for Pb, $a=0.286$~nm). 

In the paramagnetic limit the Hamiltonian can be further simplified to the form
\begin{equation}
H_e ^\sigma = -\frac{\hbar ^2}{2m} \nabla ^2 + s\mu_B H_{||} - \mu _F.
\end{equation}
If we assume the periodic boundary conditions
in the $x-y$ plane, the quasi-particle wave functions can be expressed as
\begin{equation}
\left (
\begin{array}{c}
\mathcal{U}_{k_xk_y\nu}^\sigma(\mathbf{r}) \\
\mathcal{V}_{k_xk_y\nu}^{\bar{\sigma}}(\mathbf{r})
\end{array}
\right ) = 
\frac{e^{ik_xx}}{\sqrt{L_x}} \frac{e^{ik_yy}}{\sqrt{L_y}}
\left (
\begin{array}{c}
u_{\nu} ^ \sigma(z) \\
v_{\nu} ^ {\bar{ \sigma }}(z)
\end{array}
\right )
\label{WV}
\end{equation}
where $\bar{\sigma}$ denotes the opposite spin.
In the equation (\ref{WV}) the index $i$ has been replaced by ${ k_x,k_y,\nu}$, where $k_x$, $k_y$ are the free
electron wave vector components in the $x$ and $y$ direction, while $\nu$ labels the quantum states in the
$z$ direction.
By substituting the wave function given by (\ref{WV}) into the BdG equations (\ref{BdG_start}) we obtain two 
independent set of equations
\begin{equation}
\left (
\begin{array}{cc}
H_e ^\uparrow (z)& \Delta(z) \\
\Delta(z) & -H^{\downarrow *}_e (z)
\end{array}
\right ) 
\left (
\begin{array}{c}
u_\nu ^ \uparrow(z) \\
v_\nu ^ \downarrow(z)
\end{array}
\right )=
E^{\uparrow}_\nu
\left (
\begin{array}{c}
u_\nu ^\uparrow(z) \\
v_\nu ^\downarrow(z)
\end{array}
\right ),
 \label{BdG1Da}
 \end{equation}
and
\begin{equation}
\left (
\begin{array}{cc}
H_e^\downarrow(z)& \Delta(z) \\
\Delta(z) & -H^{\uparrow *}_e(z)
\end{array}
\right ) 
\left (
\begin{array}{c}
u_\nu ^\downarrow(z) \\
-v_\nu ^\uparrow(z)
\end{array}
\right )=
E^{\downarrow}_\nu
\left (
\begin{array}{c}
u_\nu ^\downarrow(z) \\
-v_\nu ^\uparrow(z)
\end{array}
\right ),
 \label{BdG1Db}
 \end{equation}
where 
\begin{equation}
H_e^ \sigma(z)= -\frac{\hbar^2}{2m} \frac{d^2}{dz^2}+\frac{\hbar^2 k^2_{\parallel}}{2m} + s\mu_B H_{||} -\mu_F
\end{equation}
and $k^2_{\parallel}=k_x^2+k_y^2$.
Assuming that the system is infinite in the $x$ and $y$ directions ($L_x, L_y \rightarrow \infty$), 
the order parameter $\Delta(z)$ can be expressed as
\begin{eqnarray}
\Delta(z)&=&\frac{g}{2 \pi} \int d k_{\parallel} \: k_{\parallel} \sum _{\nu} \{ u_\nu ^\uparrow (z)v_\nu ^{\downarrow *}(z) \left [ 1-f(E^{\uparrow}_\nu) \right ]  \nonumber \\
&+&u_\nu ^\downarrow (z)v_\nu^{\uparrow *}(z) f(E^{\downarrow}_\nu) \}
\label{delta},
\end{eqnarray}
where $g$ is the electron-phonon coupling and $f(E)$ is the Fermi-Dirac distribution. 
The summation in Eq.~(\ref{delta}) is carried out only over these states for which the
single-electron energy $\xi _{k_xk_y\nu} ^ \sigma$
satisfies the condition  $\left | \xi _{k_xk_y \nu} ^\sigma \right | < \hbar \omega _D$,
where $\omega _D$ is the Debye frequency and $\xi _{k_xk_y\nu}^ \sigma $ is given by 
\begin{eqnarray}
 \xi _{k_xk_y\nu} ^\sigma &=& \int_0^d dz \bigg \{ u_\nu ^{\sigma *}(z)  H_e ^\sigma(z) u_\nu ^\sigma (z) \nonumber \\
 &+& v_\nu ^{\bar{\sigma} *}(z) H_e ^{\bar{\sigma}}(z) v_\nu ^{\bar{\sigma}} (z) \bigg \},
 \label{ekin}
\end{eqnarray}
where $d$ is the nanofilm thickness in the $z$-direction.
The system of equations (\ref{BdG1Da})-(\ref{BdG1Db}) and equation (\ref{delta})
are solved in a self consistent manner until the convergence is reached.
As a result the spatially varying energy gap $\Delta(z)$ is obtained. In the further analysis 
we often use spatially averaged energy gap defined as
\begin{equation}
 \bar{\Delta}=\int _0 ^d \Delta(z) dz.
\end{equation}
Since the chemical potential for nanostructures strongly deviates from the bulk
value, for each nanofilm thickness 
we determine the chemical potential by using the formula
\begin{eqnarray}
 n_e&=&\frac{1}{\pi d} \int dk_{\parallel}  k_{\parallel} \sum _{\sigma} \sum _{\nu} \int _0^d dz \bigg \{ |u_\nu^{\sigma}(z)|^2f(E_\nu)  \nonumber \\ 
 &+&|v_\nu^{\bar{\sigma}}(z)|^2 [1-f(E_\nu)]\bigg \}.
\end{eqnarray}
In the calculations we adopt the the hard-wall potential profile as the boundary conditions in the $z$ direction. \\
The set of self-consistent equations can lead to solutions with $\Delta \neq 0$ even for the values of the magnetic field 
for which the superconducting phase is already not stable - its free energy is greater than the free energy corresponding 
to the normal metal solution ($\Delta=0$). That is why in determining the critical field one should calculate and compare 
the free energy of the normal and superconducting phase, as it is done here.

\section{Results}
\label{sec:results}
In this section we analyze the superconductor to normal metal transition 
induced by the parallel magnetic field. The thickness range under consideration is chosen based on the
experiments which present measurements for the Pb nanofilms with thickness varying from 5 to 30 ML.
The calculations have been carried out for the following values of the parameters: $gN_{bulk}(0)=0.39$ where 
$N_{bulk}(0)=mk_F/(2 \pi^2 \hbar ^2)$  is the bulk density of the single-electron states at the Fermi level, 
$\hbar \omega _D=8.27$~meV, the lattice constant $a=0.286$~nm and the bulk critical temperature $T_{bulk}=7.2$~K 
which corresponds to the energy gap $\Delta_{bulk}=1.1$~meV. Within the parabolic band approximation used in the present model, 
the Fermi level in the bulk $\mu _{bulk}$ is treated as a fitting parameter.~\cite{Shanenko2007} 
Its value is determined based on the experimental results from the photoemission spectroscopy.\cite{Eom2006,Zhang2005} 
We take on $\mu _{bulk}=1$~eV which corresponds to the $\lambda _F=1.2$~nm being about four times the single monolayer
thickness $a$.

\subsection{Paramagnetic critical field oscillations}
\begin{figure}[ht]
\begin{center}
\includegraphics[scale=0.3]{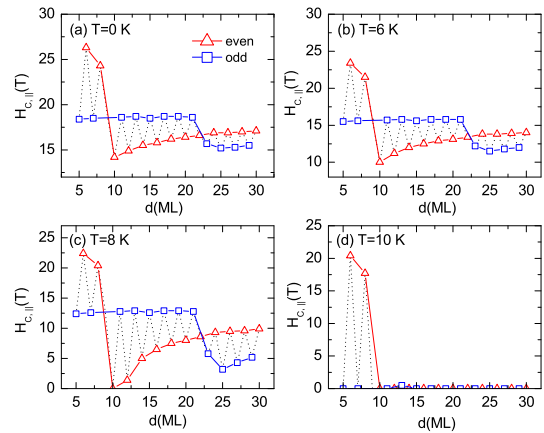}
\caption{(Color online) In-plane critical magnetic field $H_{c,||}$ as a function of the nanofilm thickness $d$ calculated for the 
temperatures $T=0,6,8$ and $10$~K.
Results for even-layered nanofilms are marked by triangular red points while for odd-layered  nanofilms by square blue points.
}
\label{fig1}
\end{center}
\end{figure}

\begin{figure}[ht]
\begin{center}
\includegraphics[scale=0.25]{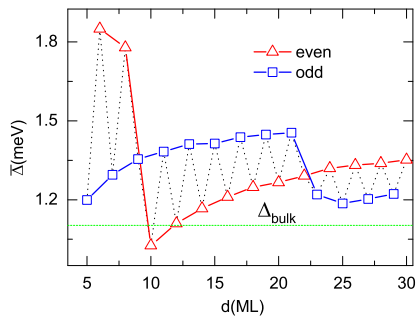}
\caption{ (Color online) Zero temperature spatially averaged energy gap $\Delta$ as a function of the nanofilm thickness $d$.
Results for even-layered nanofilms are marked by triangular red points while for odd-layered  nanofilms by square blue points.
Green horizontal line denotes the value of the energy gap  in the bulk.
}
\label{fig2}
\end{center}
\end{figure}

In Fig.~\ref{fig1} the in-plane critical field $H_{c,||}$ as a function of the nanofilm thickness is presented for different temperatures $T$. 
The value of $H_{c,||}$ is determined based on calculations of the energy gap as a function of the magnetic field and
is defined as the field for which the spatially averaged energy gap $\bar{\Delta}$ drops below $0.01\Delta _{bulk}$.
Fig.~\ref{fig1}(a) displays the oscillatory behavior of the paramagnetic critical field with the period of $2$~ML. 
In the thickness range $5-9$~ML the critical field for the even-layered films is higher than the critical field
for the odd-layered films. This relation reverses for films with the number of monolayers varying 
from 10 to 22 ML and changes again in the range $23-30$ ML. The periodicity of such modulation 
is equal to $12$~ML. The even-odd (bilayer) oscillations with the beating effect have been observed in recent measurements of 
the critical temperature~\cite{Eom2006,Guo2004} and the electron-phonon coupling in Pb thin 
nanofilms.~\cite{Zhang2010} In these experiments the modulation 
periodicity varies from 7 ML in  Ref.~\onlinecite{Eom2006} to 9 ML in Ref.~\onlinecite{Zhang2005} indicating that this quantity 
depends on the quality of the nanofilm and might vary from one experiment to another. 
Both the oscillations of $H_{c,||}$ in Fig.~\ref{fig1} result from the confinement of the electron motion in the 
direction perpendicular to the film. In the ultra thin nanofilms, the confinement of the electron motion leads to the quantization of its energy.
The Fermi sphere transforms into the series of parabolic subbands, positions of which on the energy 
scale decreases with increasing nanofilm thickness. Since the Cooper-pairing in the phonon-mediated superconductor is determined by the 
number of the electron states in the energy window $\left [ \mu - \hbar \omega _D,  \mu + \hbar \omega _D \right ]$ 
($\hbar \omega _D$ is Debye energy), the superconducting energy gap increases each time, the subsequent subband passes 
through the Fermi surface. The presented mechanism, predicted theoretically in 1963 by Blatt and Thomson~\cite{Blatt1963}, results in the tooth-like 
oscillations of the spatially averaged energy gap $\bar{\Delta}$ as a function of the nanofilm thickness. The tooth-shape of the energy gap oscillations is directly related 
to the  changes of the electron density of states in the vicinity of the Fermi level. It abruptly increases when the subband minima reach the 
Fermi level and then exponentially decreases with increasing nanofilm thickness.~\cite{Blatt1963} 
The zero temperature energy gap $\bar{\Delta}$ as a function of the number of monolayers for Pb nanofilms is presented in Fig.~\ref{fig2}.
We see that the shape of $H_{c,||}$ oscillations at $T=0$ K [Fig.~\ref{fig1}(a)] is directly related to the shape of $\bar{\Delta}$ 
oscillations in Fig.~\ref{fig2}.
It is because, in the ultra thin nanofilms with in-plane magnetic field, the pair-breaking mechanism is mainly governed by the paramagnetic (Pauli)
limit which is given by Clogston-Chandrasekhar~\cite{Clogston1962,Chandrasekhar1962} (CC) formula, $H^{CC}=\Delta_{bulk}/(\sqrt{2}\mu_B)$.
Therefore, the bilayer oscillations of 
the in-plane critical field result from the contribution of the subsequent subbands to the superconducting energy gap. 
\begin{figure}[ht]
\begin{center}
\includegraphics[scale=0.35]{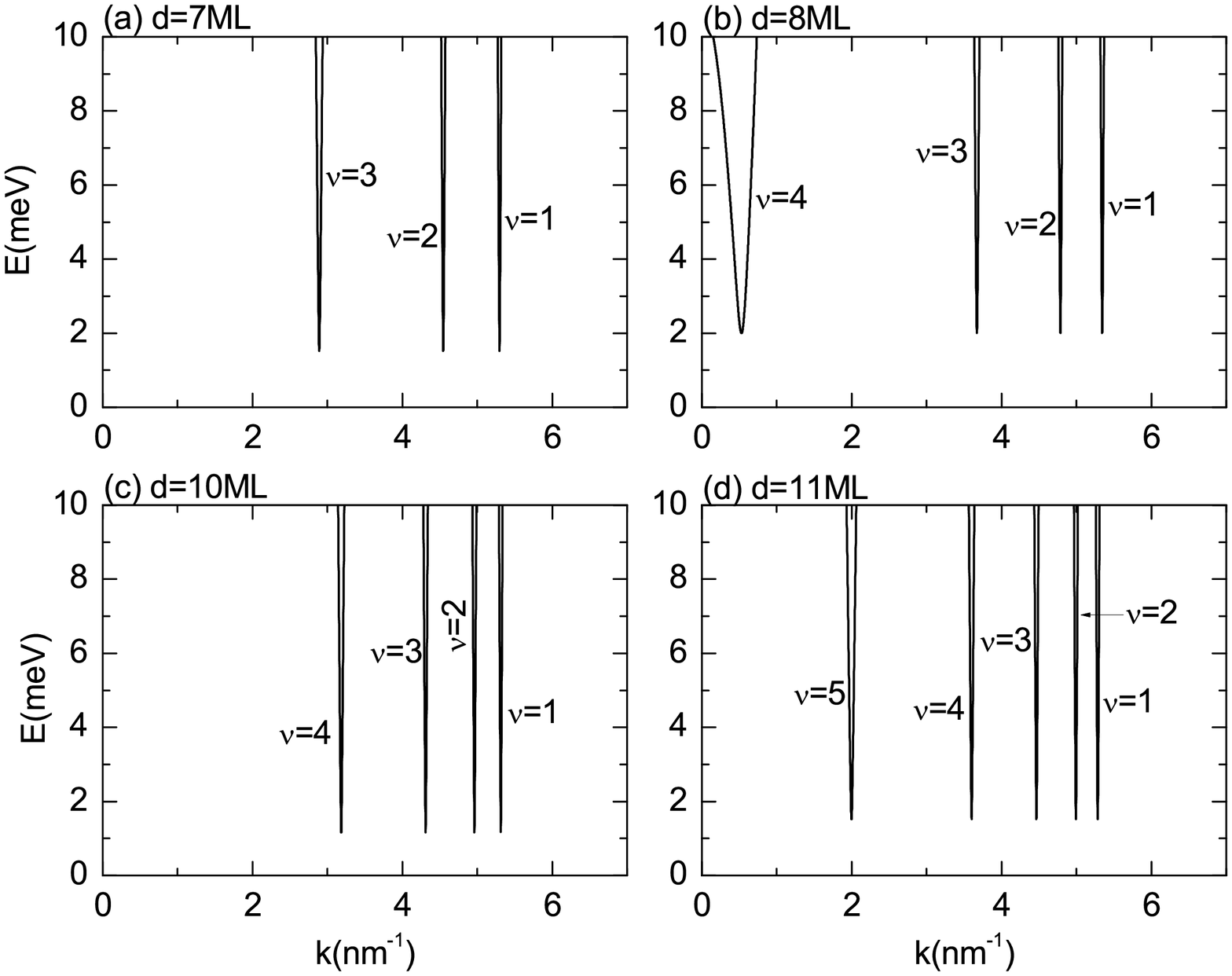}
\caption{Quasi-particle energy $E$ as a function of the wave vector $k$ for the film thickness (a) $d=7$~ML, (b) $d=8$~ML, (c) $d=10$~ML,
(d) $d=11$~ML.
}
\label{fig3}
\end{center}
\end{figure}
In Fig.~\ref{fig3} the quasi-particle energy $E$ versus the wave vector $k$ is presented for the film thickness $d=7,8$~ML (a,b) and $d=10,11$~ML (c,d). The first pair 
of thicknesses correspond to the range in which $H_{c,||}$ is higher for the even-layered films
while the second pair is related to the range for which a reversed situation is observed. We see that for the film thickness $d=7$~ML three lowest
subbands participate in the creation of the superconducting state [Fig.~\ref{fig3}(a)]. Increasing the thickness by one monolayer causes that the forth 
subband begins to contribute to the superconducting phase which leads to the enhancement of the energy gap and the critical field depicted in Fig.~\ref{fig1}(a). 
In the thickness range $5-10$~ML the subsequent subbands pass
through the Fermi level only if the number of monolayers is even. 
The period of the passages can be well estimated by the single-electron energy level  
in the form $E \approx \hbar ^2 \pi ^2 \nu^2 /(2md^2)$, where the hard-wall potential is assumed in the $z$ direction.
It allows us to determine the period $\Delta d =  \pi \hbar / \sqrt{2m \mu}$, which
for the assumed effective Fermi level gives $\Delta d=2.14$~ML. 
The difference between 2~ML and the real period of the quantum well states which pass through the Fermi surface ($\Delta d=2.14$~ML)
results in the beating effect observed in Fig.~\ref{fig1}(a). In Fig.~\ref{fig3}(c,d) 
the new subband ($\nu=5$) appears for the odd-layered film ($d=11$~ML) which is in contradiction to the pair $d=7,8$~ML [see Fig.~\ref{fig3}(a)(b)] for which 
the subsequent subband ($\nu=4$) appears for the even-layered film ($d=8$~ML). 
The interval in which  the critical field for the odd-layered films is higher than for the even-layered films ranges from $10$~ML to $22$~ML.
The source of the beating effect is clearly visible in Fig.~\ref{fig4} in which the energy of the quantum well states as a function of the nanofilm thickness 
is presented.
\begin{figure}[ht]
\begin{center}
\includegraphics[scale=0.25]{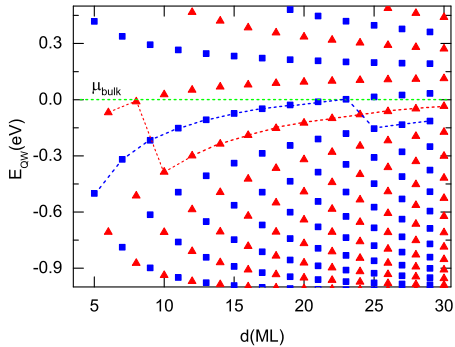}
\caption{(Color online) Energy of the quantum well states $E_{QW}$ as a function of the film thickness $d$.
Results for even-layered nanofilms are marked by triangular red points while for odd-layered  nanofilms by square blue points.
The Fermi level $\mu$ is set to zero. The energy closest to the Fermi level are joined by the blue(red) line for the even(odd) number of monolayers. 
}
\label{fig4}
\end{center}
\end{figure}
In Fig.~\ref{fig4} only the states with the energy below Fermi level (set as zero energy) are occupied and contribute to the Coooper pair condensation.
The closer to the Fermi energy they are located the larger is their contribution to the density of states. In Fig.~\ref{fig4} 
the energies closest to the Fermi level are joined by the blue (red) line for the even (odd) number of monolayers. Note that such procedure leads to the 
``structure'', in the sense of the shape, exactly the same as depicted in Fig.~\ref{fig1}(a).

Let us now discuss the value of the paramagnetic critical field obtained from our calculations. It is well-known that
in the ultra thin films with the in-plane magnetic field, the orbital-magnetic interaction is strongly reduced and the upper limit for the critical field 
is determined by the paramagnetic breakdown of the Cooper pairs. The formula for the paramagnetic critical field in the bulk 
was derived by Clogston and Chandrasekhar~\cite{Clogston1962,Chandrasekhar1962} and has the form $H^{CC}=\Delta_{bulk} / (\sqrt{2} \mu _B)$
which for Pb gives $H^{CC}=13.4$~T. This value strongly differs from the critical field obtained in our calculations which vary from $18$~T to $26$~T.
Such deviation for the Pb nanofilms has been recently reported in Ref.~\onlinecite{Sekihara2013} in which the experimental critical field $H_{C,||}$, 
has been much higher than the paramagnetic limit $H^{CC}$. It seems that such discrepancy results from the fact that due to the electron confinement 
in the nanofilm, the energy gap is enhanced (Fig.~\ref{fig2}) which leads to increase of the $H_{C,||}$.
Nevertheless, the use of the CC formula with the spatially averaged energy gap calculated for each nanofilm thickness
$H_d^{CC}=\bar{\Delta}(d) / (\sqrt{2} \mu _B)$ still produces the value of the critical field which is lower than the results 
from numerical solution of the spin-generalized BdG equations, i.e. for $d=6$~ML, $H_d^{CC}=22.7$~T while calculated $H_{C,||}=26.3$~T.
This fact can be understood if we realize that 
the superconducting energy gap depends on the $z$ coordinate and is not uniform as in the expression for
Clogston-Chandrasekhar field. Taking into account the correction which results from the averaging of the energy gap
over the $z$ coordinate we introduce the thickness-dependent parameter $\alpha(d)$ and express the critical 
field in the form
\begin{equation}
 H_{C,||}=\alpha(d) \frac{\bar{\Delta}(d)}{\mu _B}.
\end{equation}
\begin{figure}[ht]
\begin{center}
\includegraphics[scale=0.3]{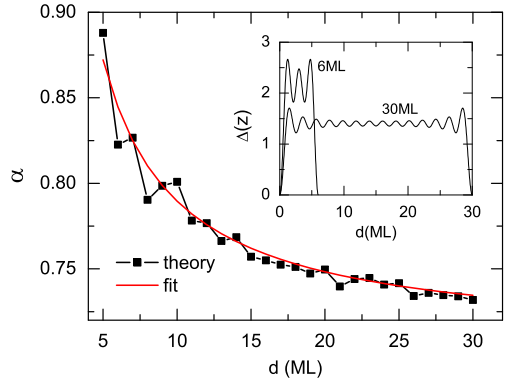}
\caption{(Color online) $\alpha$ as a function of the film thickness $d$. The red line is the fitting curve in the form $\alpha=A/d+1/\sqrt{2}$;
$A=1.17$. The inset: the spatially varying energy gap $\Delta(z)$ calculated for the nanofilm thickness $d=6$~ML and $d=30$~ML.
}
\label{fig5}
\end{center}
\end{figure}
Fig.~\ref{fig5} shows that the parameter $\alpha$ is the decreasing function of the thickness $d$ and for sufficiently
large thickness reaches the value $\alpha=0.707$. The asymptotic behavior of $\alpha (d)$ for $d \rightarrow \infty$  
results from the fact that for the large thickness the spatially varying energy gap diverges to the uniform one -
compare $\Delta(z)$ for the thicknesses $d=6$ and $30$~ML presented in the inset of Fig.~\ref{fig5}.
The dependence $\alpha(d)$ can be well fitted using the formula $\alpha=A/d+1/\sqrt{2}$ (the thickness $d$ 
denotes the number of monolayers - see Fig.~\ref{fig5}) which clearly indicates that the parameter $\alpha$  diverges to 
the bulk value $1/\sqrt{2}$ in accordance to the CC theory.
It should be noted that the smallest difference between the quasiparticle energy in the superconducting state and the Fermi energy 
is the same for each band created due to the confinement of electrons in the direction perpendicular to the sample (c.f. Fig.~\ref{fig3}). 
Using this energy gap, instead of the averaged energy gap $\bar{\Delta}$ in the CC formula, one obtains the same values of critical 
fields as the ones resulting from the BdG equations.

\subsection{Thermal effect}
In the present subsection we discuss in detail the effect of temperature on the superconductor to normal metal transition 
induced by the in-plane magnetic field. The critical field $H_{c,||}(d)$ presented in Fig.~\ref{fig1} for different 
temperatures $T$ shows that its value gradually decreases with increasing temperature. Since the critical temperature 
oscillates as a function of the nanofilm thickness (similarly to the energy gap depicted in Fig.~\ref{fig2}), at the temperature 
$T=10$~K, the nanofilms with the thickness $d=7,8$~ML are superconducting while the film with the thickness 
therebetween transits to metallic state. 

\begin{figure}[ht]
\begin{center}
\includegraphics[scale=0.35]{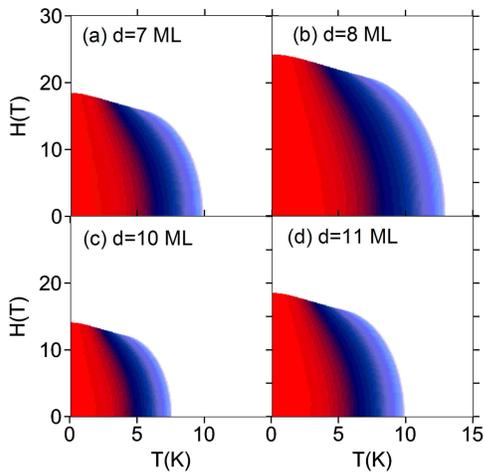}
\caption{(Color online) Spatially averaged energy gap $\bar{\Delta}$ as a function of the magnetic field $H$ and temperature $T$
for the nanofilm thickness (a) $d=7$~ML, (b) $d=8$~ML, (c) $d=10$~ML and (d) $d=11$~ML. The value of energy gap in each figure is 
normalized with respect to its maximum.
}
\label{fig6}
\end{center}
\end{figure}

\begin{figure}[ht]
\begin{center}
\includegraphics[scale=0.25]{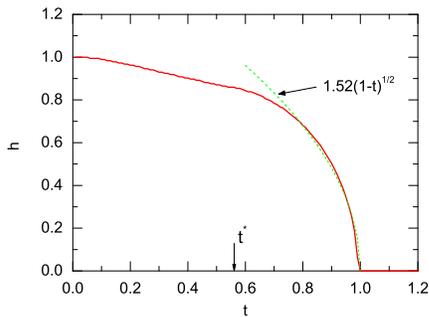}
\caption{(Color online) Critical field $h=H_{c,||}/H_{c,||}(0)$ as a function of $t=T/T_c(0)$ obtained from the solution of BdG equations.
Here the dashed green line corresponds to $h=1.52(1-t)^{1/2}$. The first to second order transition, $t^{*}$, is marked by arrow.
}
\label{fig7}
\end{center}
\end{figure}

\begin{figure}[ht]
\begin{center}
\includegraphics[scale=0.25]{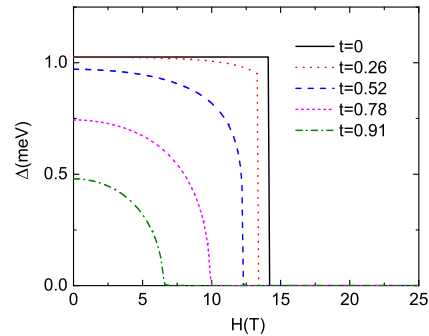}
\caption{(Color online) Spatially averaged energy gap $\bar{\Delta}$ as a function of magnetic field $H$ for different temperatures. The first to second
order transition emergences at $t=t^*=0.56$. Calculations performed for the film thickness $d=10$~ML.
}
\label{fig8}
\end{center}
\end{figure}

In Fig.~\ref{fig6} we present the spatially 
averaged energy gap as a function of the magnetic field and temperature. The results in Fig.~\ref{fig6} have been calculated for the 
nanofilm thicknesses $d=7,8,10$ and $11$~ML. To make it more transparent, the value of the energy gap in each of the figures is 
normalized with respect to its maximum.
In Fig.~\ref{fig6} we can see that the range of the magnetic field and 
temperature for which the sample is superconducting  is larger for the thickness for which the enhancement of the 
energy gap (see Fig.~\ref{fig2}) is found. Nevertheless the thickness-dependence is not visible in
the normalized critical magnetic field $h=H_{c,||}/H_{c,||}(0)$ as a function of $t=T/T_c(0)$, where $H_{c,||}(0)$ and $T_c(0)$ is 
the critical field at $T=0$ and the critical temperature for $H=0$, respectively. 
In other words the phase diagram $h-t$ presented in Fig.~\ref{fig7} looks the same for each film thickness which 
is understandable if we recall our restriction to the Pauli limit with no orbital effect.
The initial behavior of $h$ at the temperature close to $T_c$ ($t=1$) can be well estimated by the formula 
$h=A(1-t)^{1/2}$. Note that, in contrast to the orbital limit with no Pauli effect,
the slope of $h$ at critical temperature $t=1$ is infinite and in the vicinity of $t=1$ is expressed by $dh / dt \propto \sqrt{1/(1-t)}$.
In this range the transition at $H_{c,||}$ is of second order. In the Pauli limit the second order transition is suppressed with 
decreasing temperature and below $t^*=0.56$ it becomes the first order.~\cite{Maki1964,Maki1966} The first to second order transition 
is clearly presented in Fig.~\ref{fig8} in which the energy gap versus magnetic field is presented for different 
temperatures. 
Note that the critical field $H_{c,||}$ at $T=0$ is not equal to the paramagnetic limit $H^{CC}$ but is further enhanced. 
A detailed discussion of the deviation of the critical field from $H^{CC}$ is presented in  sec.~\ref{sec:results}.A.

\subsection{Thickness dependent electron-phonon coupling}
The microscopic model based on the spin-generalized BdG equations 
allows to analyze superconductor to normal metal transition in the nanofilms when its thickness is reduced to few nanometers.
However, in the nanoscale regime, the confinement affects not only the 
electronic spectrum, but also the phononic degrees of freedom. The phonon dispersion in thin films strongly deviates 
from that in the bulk.~\cite{Nabity1992}
The quantization of the phononic spectra in the nanofilms and its influence on $T_c$ and energy gap oscillations have
been considered in Ref.~\onlinecite{Hwang2000}. However, the effect of the confinement on the 
electron-phonon coupling strength has not been included. This lack has been recently supplemented 
by Saniz et al. in Ref.~\onlinecite{Saniz2013}. In this paper~\cite{Saniz2013} the authors have investigated the 
effect of confinement on the strength of the 
electron-phonon coupling as well as the electronic spectrum and its influence on 
the oscillations of the critical temperature.  
The formula for the phonon-mediated 
attractive electron-electron interaction has been derived with the use of the Green function approach beyond the contact potential approximation. 
It has been found that the increase of the critical temperature observed in superconducting nanofilms 
is due to the increase of the number of phonon modes what results in the enhancement of the electron-phonon coupling.  
In contradiction to previous models~\cite{Shanenko2006, Shanenko2007}, this study predicts the suppression of the critical 
temperature with increasing density of states at the Fermi level. Such conclusion seems not to be confirmed
by recent experiments in which the direct correlation between the $T_c$ oscillations and the energy distribution of the quantum 
well states in nanofilms is observed.~\cite{Eom2006} 
\begin{figure*}[hbt]
\begin{center}
\includegraphics[scale=0.18]{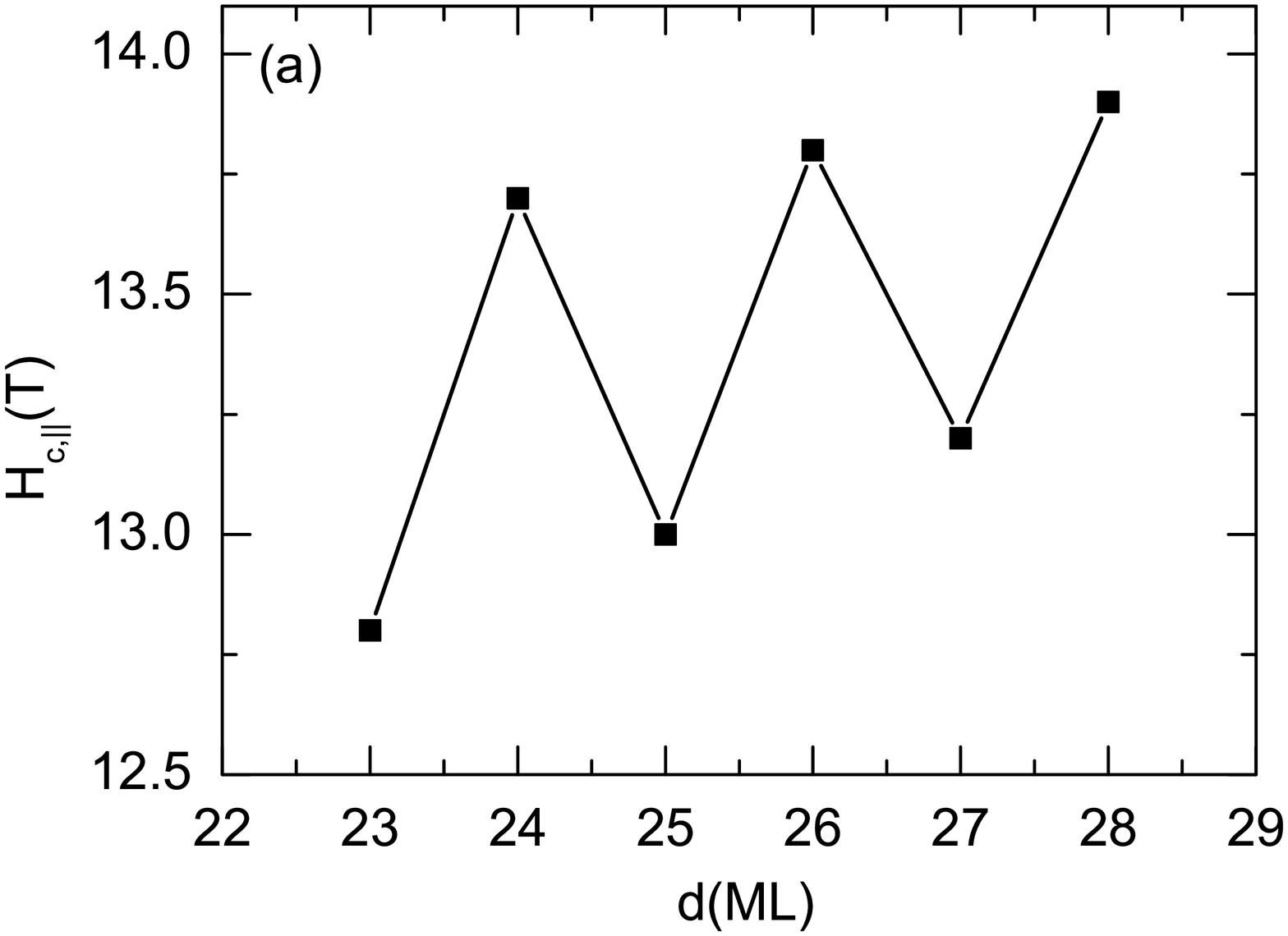}
\includegraphics[scale=0.18]{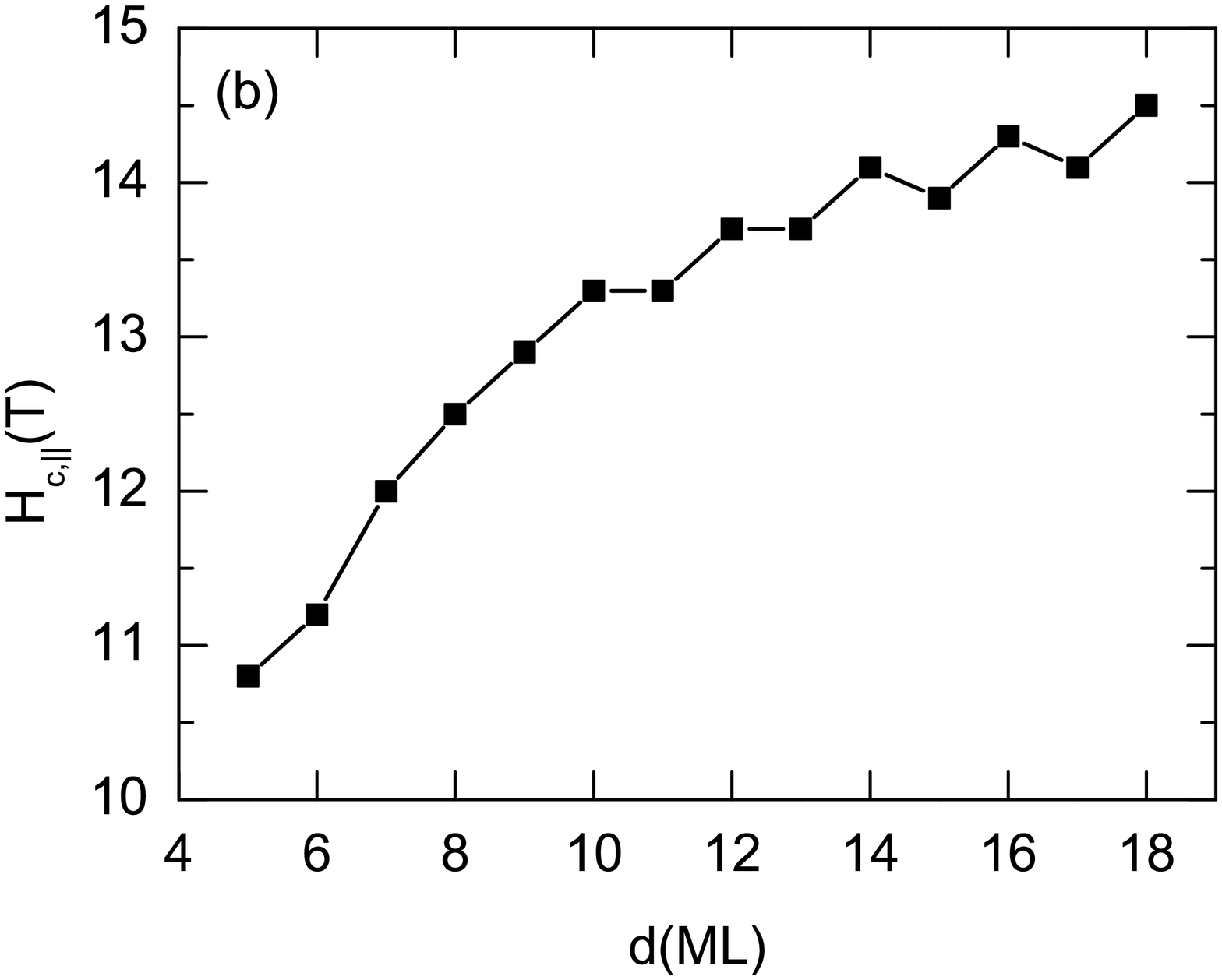}
\includegraphics[scale=0.18]{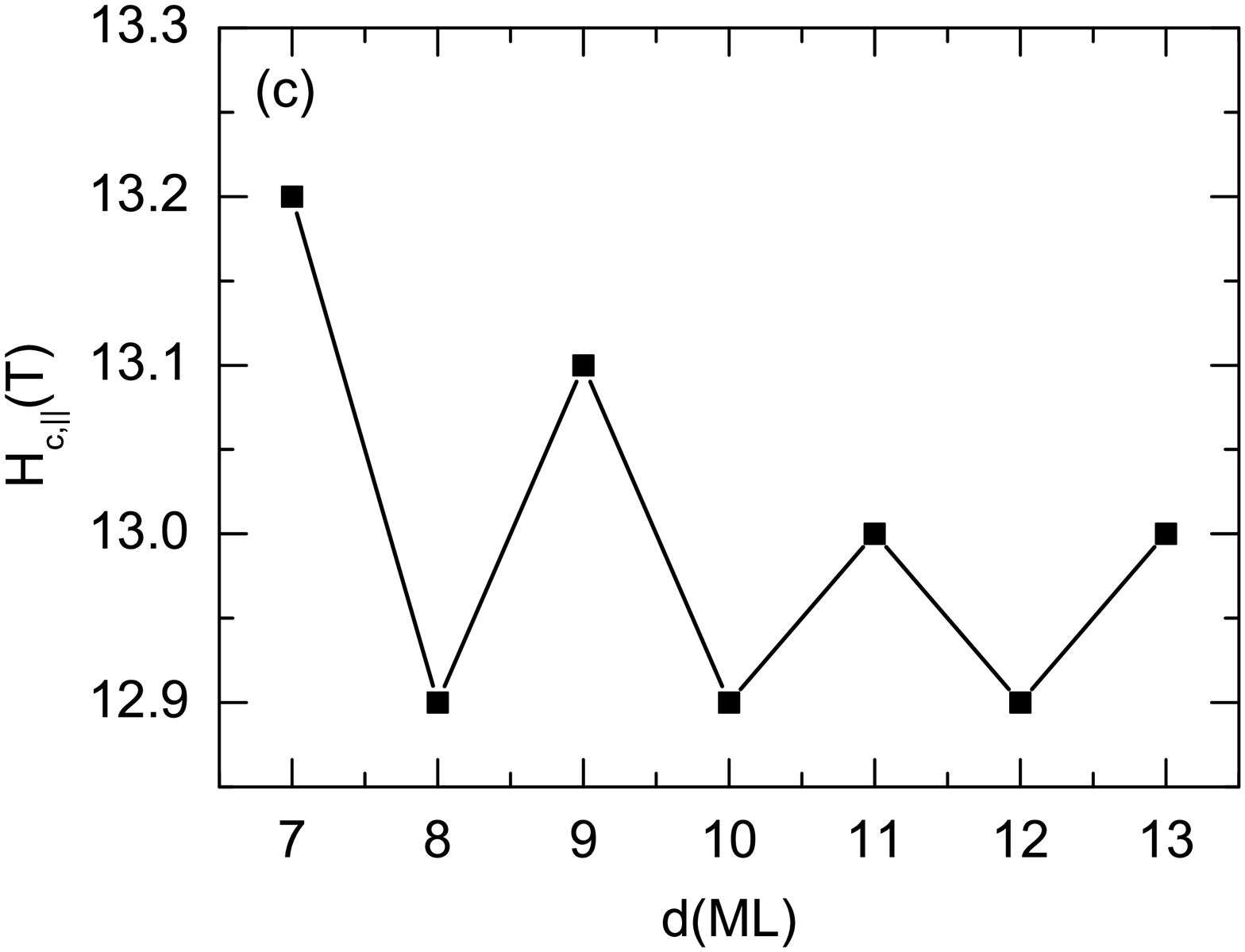}
\caption{Zero temperature critical field as a function of the nanofilm thickness. Results calculated 
for the parameters determined from the fitting of $T_c$ with experimental measurements from
(a) Ref.~\onlinecite{Brun2009}, (b) Ref.~\onlinecite{Ozer2006} and (c) Ref.~\onlinecite{Eom2006} (see table~\ref{tab1}).
} 
\label{fig9}
\end{center}
\end{figure*}
In order to determine the realistic value of the electron-phonon coupling which occurs in experiments, there is one more aspect which 
should be taken into consideration. Note that the results presented in this paper regard the so-called freestanding nanofilms. 
In fact, Pb films are grown on Si(111) substrate and, for the purpose of the transport measurements, are covered by the layer of Au. 
Both these layers strongly affect the electron-phonon coupling in the nanofilm due to the interface effect.
It has been recently observed~\cite{Zhang2005} that the electron-phonon coupling for Pb nanofilms on Si substrate 
is lower than in the bulk and diverges to the bulk value with increasing film thickness. 
The results presented in Ref.~\onlinecite{Zhang2005} indicate that the interface effect
has a strong impact on the electron-phonon coupling and should be taken into account in the presented model in order to 
reproduce the experimental data. The simple approximation which allows to take this effect into account  has been 
recently proposed by Chen et al. in Ref.~\onlinecite{Chen2013}. It has been assumed~\cite{Chen2013} 
that in the first approximation the spatially-dependent electron-phonon coupling can be well estimated by the formula
\begin{equation}
 g(z)=
 \left \{
 \begin{array}{cc}
  g_{if} & 0 < z < d_{if} \\
  g_0 & d_{if} < z < d \\
 \end{array}
\right . ,
\end{equation}
where $d_{if}$ is the interface thickness with the electron-phonon coupling $g_{if}$ and $g_0$ is the electron-phonon coupling in the bulk.
Averaging over the $z$ coordinate and including  the oscillatory behavior of $g$ coefficient reported in experiments~\cite{Zhang2005}
leads to the formula (for details see Ref.~\onlinecite{Chen2013})
\begin{equation}
 \bar{g}=g_0-\frac{g_1 \left ( \frac{4 \pi a N}{\lambda _F} \right ) } { N } ,
\end{equation}
where $g_1(x)$ is a periodic function, $N$ is the number of monolayers, $\lambda _F$ is the Fermi wavelength and $a$ is the lattice constant. 
Since the period of $T_c$ oscillations in Pb nanofilms is measured to be $\sim 2$~ML, the authors of Ref.~\onlinecite{Chen2013} have reduced
the function $g_1(x)$ to only two parameters: $g_1(\pi)$ for an odd number of monolayers and $g_1(2 \pi)$ for an even number of monolayers.
By appropriate choice of these parameters and by setting the Fermi level to the value corresponding to the period of the quantum-size oscillations ($2$~ML)
one can well reproduce the experimental data of $T_c(d)$ coming form Ref.~\onlinecite{Ozer2006,Eom2006,Brun2009}.
The values of the parameters used for each of the mentioned references is presented in table~\ref{tab1}.
\begingroup
\squeezetable
\begin{table}
\begin{ruledtabular}
\begin{tabular}{cccc}
Reference & $g_0N(0)$ & $g_1(\pi)$ & $g_1(2\pi)$ \\
\hline\hline
Ref.\onlinecite{Eom2006} & 0.36 & $-0.12g_0$ & $0.84g_0$ \\
Ref.\onlinecite{Ozer2006} & 0.39 & $0.64g_0$ & $1.46g_0$ \\
Ref.\onlinecite{Brun2009} & 0.39 & $1.67g_0$ & $2.13g_0$ 
\end{tabular}
\end{ruledtabular}
\caption{The value of parameters used in Ref.~\onlinecite{Chen2013} to reproduce the oscillations of $T_c$ observed in experiments 
- see Reference column.}
\label{tab1}
\end{table}
\endgroup
In the present subsection we use these parameters to predict the zero temperature
in-plane critical field as a function of the nanofilm thickness. We believe that such procedure allows to determine the realistic 
value of the critical field comparable with recent experimental measurements.~\cite{Sekihara2013}
The thickness-dependent oscillations of the zero temperature critical field are presented in Fig.~\ref{fig9}.
Although the periodicity of the $H_{c,||}$ oscillations for each of the considered references is the same and equal ~2 ML,
the characteristics of these oscillations are different and vary form one experiment to another. This difference regards 
the overall trend of $H_{c,||}$ which is an increasing function of the thickness in case of Refs.~\onlinecite{Brun2009, Ozer2006}
but decreasing in the case of Ref.~\onlinecite{Eom2006}. Note that the values of the critical field in Fig.~\ref{fig9} are suppressed as compared 
to the critical field calculated with the electron-phonon coupling for the bulk [compare with Fig.~\ref{fig1}(a)].

\section{Summary}
\label{sec:concl}
The superconductor to normal metal phase transition driven by the magnetic field for Pb nanofilms is investigated with the 
use of the BdG equations. Only the Pauli pair-breaking mechanism is included as it is assumed that the external magnetic 
field is parallel to the surface of the nanofilm.  It is shown that the even-odd oscillations of the critical magnetic field 
appear as the thickness of the sample is increased. The shape of the $H_{c,||}$ oscillations is directly related to the shape of 
the thickness dependence of the superconducting gap. The beating effect visible in the oscillations is discussed in 
the context of the energy of the quantum well states which appears due to the confinement of the electron 
motion in the direction perpendicular to the sample. As it has been shown the period in the nanofilm thickness between two neighbouring 
peaks in the critical field is equal to 2.14 ML. 
As the number of the monolayers has to be an integer the beating effect appears in the oscillatory behavior 
of $H_{c,||}$ and $\bar{\Delta}$. We also show that the zero-temperature critical field in the nanofilms is higher than  the 
Clogston - Chandrasekhar paramagnetic limit and it diverges to the CC value  for sufficiently thick films.
This fact is explained in term of the spatially varying energy gap induced by the confinement.
The phase diagrams in the $(H,T)$ plane are presented for different values of $d$. 
According to the obtained results the thickness-dependence is not visible in the phase diagrams when both the field and the 
temperature are  normalized $h=H_{c||}/H_{c||}(0)$ and $t=T/T_c(0)$. Moreover, the interface effect on the  electron-phonon 
coupling is included by using a simple approximation proposed in Ref.~\onlinecite{Chen2013}. This leads to a spatially dependent $g$ factor and its oscillatory 
behavior with increasing nanofilm thickness. As it is shown this effect leads to a suppression of the critical field values in comparison 
to the ones corresponding to the bulk. We believe that such approach allows for comparison of the calculated 
$H_{c,||}$ oscillations with the experimental results.

\section{Acknowledgments}
Discussions with J{\'o}zef Spa{\l}ek are gratefully acknowledged.
This work  was financed from the budget for Polish Science in the years 2013-2014. Project number: IP2012 048572.
M. Z. acknowledges the financial support from the Foundation for Polish Science (FNP) within project TEAM.


%

\end{document}